\documentclass[12pt,a4paper]{article}
\newcommand{\be}{\begin{equation}}
\newcommand{\ee}{\end{equation}}
\newcommand{\bea}{\begin{eqnarray}}
\newcommand{\eea}{\end{eqnarray}}
\newcommand{\ep}{\epsilon}

\begin{document}

\begin{titlepage} 

\begin{centering} 
{\bf {D-particle Recoil Space Times and ``Glueball'' Masses}} 

\vspace{0.5in} 

{\bf {Nick E. Mavromatos}} \\

{\it {Department of Physics, Theoretical Physics, King's College London,
The Strand, London WC2R 2LS, United Kingdom,}} \\

\vspace{0.1in}
and \\
\vspace{0.1in} 
{\bf {Elizabeth Winstanley}} 

{\it {Department of Physics, Theoretical Physics, University of Oxford, 
1 Keble Road Oxford OX1 3NP, United Kingdom.}} \\

\vspace{1in}

{\bf {Abstract}} 

\end{centering} 

{\small {We discuss the properties of matter in a $D$-dimensional 
anti-de-Sitter-type space time 
induced dynamically by the recoil of a very heavy D(irichlet)-particle 
defect embedded in it. 
The particular form of the recoil geometry,
which from a world-sheet view point follows from 
logarithmic conformal
field theory deformations of the pertinent sigma-models, 
results in the presence of {\it {both}} 
infrared and ultraviolet (spatial) cut-offs.
These are crucial in ensuring the presence of mass gaps
in scalar matter propagating in the D-particle recoil space time.  
The analogy of this problem with the Liouville-string approach to QCD,
suggested earlier by John Ellis and one of the present authors, 
prompts us to identify the resulting scalar masses with those 
obtained in the supergravity approach based on the Maldacena's 
conjecture, but without the imposition of any supersymmetry in our case. 
Within reasonable numerical uncertainties, we observe that agreement 
is obtained between the two approaches for a particular value 
of the ratio of the two cut-offs of the recoil geometry.
Notably, our approach does not suffer from the ambiguities of the 
supergravity approach as regards the validity of the comparison of
the glueball masses computed there with those obtained 
in the continuum limit of lattice gauge theories.}}

\end{titlepage} 

\section{Introduction and Summary}

The conjecture of Maldacena~\cite{maldacena} 
that (quantum) correlators on large-N supersymmetric 
gauge-field theories, living on a space-time manifold ${\cal {M}}$,  
may be computed by purely classical 
supergravity methods in the bulk of an Anti-de-Sitter (AdS) 
space-time with ${\cal {M}}$ as its boundary, found 
a very interesting application to Quantum Chromodynamics (QCD),
after Witten's suggestion~\cite{witten} that such an approach 
may lead  
to an understanding of the 
confinement-deconfinement transition. 
This created an enormous interest in the subject~\cite{oz}.

However, most of the approaches so far have used the 
original formulation, based on critical (super)string theory,
which at a certain stage makes 
explicit use of space-time supersymmetry. The latter is
broken explicitly by ``temperature'' in the approach of \cite{witten}, 
in order to provide a regularized version of zero-temperature QCD.  
In addition, there appear to be some doubts that the calculations 
of glueball masses 
based on supergravity calculations in the 
bulk of AdS geometry~\cite{koch}-\cite{zyskin}, 
which, from a superstring-theory viewpoint, 
are lowest order in $\alpha'$,
are strictly applicable to the case of glueball 
masses obtained 
in the continuum limit of lattice QCD computations~\cite{lattice}, 
which seem to necessitate a resummation of higher order corrections
in $\alpha '$. The latter are still uknown in the context of 
string theory, although there are claims~\cite{csaki} that such 
corrections do not affect much the results on the ratios of the 
glueball masses
calculated by means of the supergravity approach.

An alternative to the above-described critical-dimension 
superstring theory approach, which necessarily 
makes use of space-time supersymmetry 
at a certain stage, 
is the Liouville string approach to 
QCD~\cite{liouvilleqcd}, where also
AdS space times appear but supersymmetry is not a requirement.
In this context, but in a different approach from \cite{liouvilleqcd}, 
with emphasis on the r\^ole of magnetic monopoles 
on the confinement problem, 
John Ellis and one of the authors 
have argued~\cite{em} that AdS space time 
appears naturally and {\it {dynamically}} 
within the modern context of 
D-brane approach to gauge theories. This is a result of 
quantum fluctuations of the monopole defect, described by ``recoil'' 
logarithmic conformal 
field theories in a world-sheet non-critical (and non-supersymmetric)
string approach~\cite{kogan}. 
More specifically, the D-particle recoil model of
\cite{kanti} has been used in the limit $u_{i}\rightarrow 0$, which,
arguably, describes the interaction of a quark-antiquark
Wilson loop with a magnetic monopole for the gauge field
configuration in QCD~\cite{em}.
Such configurations have been argued by Polyakov \cite{polyakov}
and 't Hooft \cite{thooft} to play a crucial r\^{o}le in the 
confinement problem.

What we have argued in \cite{em}
is that the fluctuating monopole 
may be represented, in this non-critical string framework,
as a quantum fluctuating D(irichlet)-particle.
Such fluctuations may be considered as a specific case of ``recoil'',
which in turn,
corresponds to logarithmic worldsheet 
deformations~\cite{kogan} of the pertinent $\sigma $-model.
The Wilson loop in this picture corresponds to a static macroscopic
closed string in interaction with the D-particle defect.
Due to the static nature of the loop, the recoil 
velocity $u_{i}$ of the defect 
is taken to be {\it {vanishing}}. In the formalism of 
logarithmic conformal field theory of \cite{kogan}, 
this limit corresponds to considering only the effects of 
one of the deformations of the logarithmic pair, namely the C-deformation. 

In reference \cite{em}, we did not present 
a complete mathematical analysis of all the properties of the 
induced recoil space time, and in particular we did not address the question
whether, in this formalism, there is a gap in the associated 
glueball spectrum, according to the AdS/conformal-field theory 
correspondence~\cite{witten}.
This is the point of the present article. However, 
we shall follow a slightly 
different approach than that in ref. \cite{em}. 
Here, as in ref. \cite{em}
and \cite{kanti}, 
the Liouville field is 
identified with the target time which, however, 
we shall not consider it as constant, in contrast
to the situation in \cite{em}. Instead 
we shall fix a specific 
combination of time and spatial co-ordinates. 
As we shall see, this yields a mass gap in the spectrum 
of scalar matter propagating in the induced space time,
and hence is appropriate for a discussion of glueballs in 
(zero-temperature) QCD. 
However, the basic 
philosophy of using D-particle models 
as regulators of QCD, which underlies 
both approaches, remains identical between the two 
scenaria.  

It is important to note, 
that in our approach and that of ref.~\cite{em} 
there is no need for a large-N
assumption as in the 
supergravity approach~\cite{maldacena,witten}; 
the large AdS radius, and the associated small Regge slope 
$\alpha '$,  
needed for the validity of the 
low-energy supergravity Lagrangian in the bulk of AdS space, 
where one can calculate
reliably, arise naturally~\cite{em}. 
In particular, the radius of the induced AdS space time
is found to be proportional to $|\epsilon |^{-2}$, 
where $\epsilon $
is the standard regulating parameter of the logarithmic recoil
operators \cite{kogan}, assumed small for the validity of the 
world-sheet analysis. Closure of the logarithmic world-sheet 
algebra~\cite{kogan} requires that $|\epsilon|^{-2} \sim {\rm {ln}}A$,
where $A$ is the area of the Wilson loop, which in the infrared
regime, in which we are interested, is assumed large~\cite{em}.
Thus in the infrared regime of QCD, the AdS radius 
is large, independently of the number of colours
$N$ of the gauge group, in contrast to the supergravity approach where 
the AdS squared radius is proportional to ${\sqrt{g_{s}N}}$,  
with the string coupling being 
connected to the gauge coupling as $g_{s}=g_{YM}^2$.  

More importantly, it seems to us that 
the calculated glueball masses in our approach 
are calculated in the same limit~\cite{em} 
as the
glueball masses in the continuum limit of the lattice QCD 
calcuation~\cite{lattice}, 
and hence a direct comparison is possible.  
This is due to the fact that the glueball masses 
$M$ are given in terms of a fixed ultraviolet cut-off 
of the effective theory, $\Lambda _{UV}$, 
as 
\be
M^2 =f(g^2_{YM}N) \Lambda ^2_{UV}
\label{gmasses}
\ee
where $f$ is some function. In the supergravity approach
one should consider the (fixed) limit $g_{YM}^{2}N \gg 1$, which 
necessitates 
a large N approach in order to have a small string coupling 
$g_{s}=g_{YM}^{2}$. 
On the other hand, in the continuum limit of 
lattice gauge theories~\cite{lattice}, 
$\Lambda^{2}_{UV} \rightarrow 0$,
which implies $ g_{YM}^{2} N \rightarrow 0$. 
In our approach~\cite{em},  since the
radius of the AdS is proportional to $|\epsilon|^{-2}$, which is small in 
the infrared limit of QCD, independently of the number of colours,   
one can naturally achieve 
weak string couplings independently of a large N limit, and thus one
can compute glueball masses in the same limit as the continuum limit
of lattice QCD. 

As we shall see, however, under certain conditions,  
there can be agreement with the results of \cite{koch}-\cite{zyskin},
which may be taken to be an independent confirmation 
of the claims~\cite{csaki} 
that higher-order corrections 
in $\alpha '$, within the supergravity approach, do not change 
qualitatively the results, at least as far as the ratios of glueball masses
are concerned.   

In the present article  
we first describe in detail how an AdS space time arises from D-particle
recoil, as in \cite{kanti}, and subsequently we 
demonstrate the crucial r\^{o}le of the recoil picture
in inducing a gap in the scalar glueball spectrum at zero
temperatures. This arises due to the presence of {\it both}  
ultraviolet and  infrared (spatial) cut-offs in the recoil geometry, 
which appear in order to avoid curvature singularities in the
space time.
Namely, we shall show that the time variable, which in the 
present article,
as in \cite{em,kanti}, is identified with the Liouville field, 
needs to be
{\em {strictly positive}} in order for the gap to exist.
This positivity reflects the fact that we are considering the induced
space time as a result of (and hence consequently after) 
the recoil~\cite{em,kanti}.
However, for the existence of the gap in the spectrum, we need the
simultaneous presence of a second cut-off.
It is the presence of this cut-off which implies 
that the space time (for low temperatures) is inside the boundary of
AdS, set~\cite{em} by the radius $b\propto |\epsilon |^{-2}$. 

We emphasize that two cut-offs are necessary for the appearance
of a gap in the non-black hole AdS, in contrast to the case of
\cite{witten}, where the space time with a regular AdS (non-black hole)
does not lead to a gap in the glueball spectrum. 
This is the reason why Witten~\cite{witten} considered 
AdS black holes, to study the deconfinement transition in QCD starting
from the high-temperature regime of the model, where the black hole
space times are known to be thermodynamically stable \cite{hpage}. 

We stress that 
in the Liouville string case studied in  \cite{em} and here,
one is working at zero temperature. In this work we demonstrate
the existence of a glueball
mass gap for the regular non-black hole AdS,
which is the stable space time at low temperatures \cite{hpage}.
This property demonstrates the possibility of having glueballs
in our picture at zero temperature, which is the situation met
in realistic QCD models.

This picture also complements nicely the approach of \cite{amit}
where a gas of AdS black holes in various (finite) temperature regimes
has been studied.
In that work, in the high temperature phase, where the black holes
are stable, and a world-sheet perturbation theory breaks down,
it was found that the radius of AdS was small,
scaling like the inverse of the temperature.
On the other hand, at low temperatures, where there are no black
holes, 
the situation may be thought of as the limiting case of a black hole AdS
space time with vanishing black hole mass. This is the case we study here,
but we approach the problem in a straightforward way,
deriving the metric directly in the zero-temperature 
recoil picture of \cite{em}. Notably, in this regime
the world-sheet perturbative approach is reliable~\cite{em,kogan}.

The nature of the phase transition from low to high temperatures,
i.e. from confinement to deconfinement, is not completely understood as yet,
and the subject 
needs further study. However, 
the analogy with the Van der Waals case, 
pointed out in \cite{amit}, is 
suggestive of a first order transition,
which seems to be supported by the black-hole thermodynamical  
approach of refs.~\cite{witten,hpage}.

Before commencing our analysis, 
we note, for the benefit of the reader,  
that in the present article 
our convention for the signature of a Lorentzian space time metric is 
$(-++\ldots +)$,   
and we use units in which $G=\hbar =c =1$. This will be understood 
in what follows.

\section{Glueball masses in Supergravity}

Witten \cite{witten} was the first to show that there is a mass
gap, in the three-dimensional sense, for quantum fields propagating
on the five-dimensional space time 
\be
ds^{2}=\left( \frac {\rho ^{2}}{b^{2}} -
\frac {b^{2}}{\rho ^{2}} \right) d\tau ^{2}
+ \left( \frac {\rho ^{2}}{b^{2}} -
\frac {b^{2}}{\rho ^{2}} \right) ^{-1}  d\rho ^{2}
+\rho ^{2} \sum _{i=1}^{3} dx_{i}^{2}.
\label{wittenmetric}
\ee
This metric is constructed from the five-dimensional 
Euclidean Schwarzschild anti-de Sitter (AdS) geometry
\be
ds^{2}= \left( \frac {r^{2}}{b^{2}} +1 -\frac {M}{r^{2}} 
\right) dt^{2} +
\left( \frac {r^{2}}{b^{2}} +1 -\frac {M}{r^{2}} 
\right) ^{-1} dr^{2}+r^{2} d\Omega _{3}^{2}
\ee
in the limit of large mass $M$, as follows (we have set
the gravitational coupling constant equal to unity in order
to simplify the algebra). 
Firstly, the co-ordinates $r$ and $t$ are rescaled:
$r\rightarrow \left( \frac {M}{b^{2}} \right) ^{\frac {1}{4}} \rho,
~t \rightarrow \left( \frac {M}{b^{2}} \right) ^{-\frac {1}{4}} \tau
$
so that, for large $M$,
$\frac {r^{2}}{b^{2}}+1-\frac {M}{r^{2}} \rightarrow
\left( \frac {M}{b^{2}} \right) ^{\frac {1}{2}} \left[
\frac {\rho ^{2}}{b^{2}} -\frac {b^{2}}{\rho ^{2}} \right].$
Then the metric becomes
\be
ds^{2}=\left( \frac {\rho ^{2}}{b^{2}} -
\frac {b^{2}}{\rho ^{2}} \right) d\tau ^{2}
+ \left( \frac {\rho ^{2}}{b^{2}} -
\frac {b^{2}}{\rho ^{2}} \right) ^{-1}  d\rho ^{2}
+\left( \frac {M}{b^{2}} \right) ^{\frac {1}{2}}
\rho ^{2} d\Omega _{3}^{2}.
\ee
The factor multiplying the last term in this metric means that
the radius of the $S^{3}$ is of order $M^{\frac {1}{4}}$ and so
diverges as $M\rightarrow \infty $.
Therefore one can introduce local flat co-ordinates 
$x_{i}$ near a point
$P\in S^{3}$.
This gives the metric (\ref{wittenmetric}).
Witten then considers a massless scalar field on this background,
having form
\be
\Phi (\tau, \rho ,x_{i})=f(\rho )\exp \left( i
\sum _{i=1}^{3}k_{i}x_{i} \right) ,
\ee
where $f$ satisfies the equation
\be
-\frac {1}{\rho } \frac {d}{d\rho } \left[
\rho ^{3} \left( \frac {\rho ^{2}}{b^{2}} -
\frac {b^{2}}{\rho ^{2}} \right) \frac {df}{d\rho }
\right] +k^{2} f=0.
\label{witteqn}
\ee
As $\rho \rightarrow \infty $, (\ref{witteqn}) has two
linearly independent solutions, which behave as 
$f\sim {\mbox {constant}}$ and $f\sim \rho ^{-4}$.
For a normalizable solution, the latter behaviour is required.
In addition, Witten shows that there are no normalizable
solutions when $k^{2}$ is positive.
At the horizon ($\rho \rightarrow b$), Witten stated that the
required boundary condition is $\frac {df}{d\rho }\rightarrow 0$.
However, further investigation has shown that the 
appropriate boundary condition is that $f$ is regular at $\rho =b$.
It is shown in \cite{koch} that the boundary condition specified
by Witten is in fact not realized at the horizon, although 
specifying this condition just outside the horizon is in fact
the best way to proceed numerically \cite{ooguri}.
Either way, the boundary condition at the horizon will only be 
satisfied for discrete, negative values of $k^{2}$, leading to a 
discrete mass spectrum for $m^{2}=-k^{2}$ (which is the mass
in the three-dimensional sense, for the Euclidean geometry).
These discrete eigenvalues have been calculated in 
\cite{koch}-\cite{zyskin}.
Later, we shall compare the numerical eigenvalues computed
in  this approach with our predictions from the recoil 
geometry.

The conclusions of this calculation are applied to
glueball states using the AdS/CFT correspondence~\cite{maldacena}-\cite{oz}.
The mass gap for quantum fields in the
bulk of the Euclidean, asymptotically AdS, geometry
implies that there is also a mass gap for glueball
states on the boundary of AdS. 
In this paper we shall only discuss the massless scalar
field equation, from which one can conclude, via the 
AdS/CFT correspondence, information about the
masses of the $O^{++}$ glueball states.
Other glueball states may be studied by using
other quantum field equations, for example, 
information about the $O^{--}$ glueball states 
can be gleaned from the equation for a two-form
field in the bulk \cite{csaki}.
Here we consider only the simplest case, in order
to compare our approach with the study
of glueball masses from supergravity. We hope to come back to a
more systematic analysis of the higher-rank 
glueball states in a future publication. 

Before we consider the mass gap produced on the recoil 
space time, which will be discussed in the next section, 
we first review why there is no
mass gap on the (Euclidean AdS) ball, in order to emphasize the
physical differences between the two geometries.
Consider $D$-dimensional Euclidean AdS with metric
\be
ds^{2}=\frac {|\ep |^{-8}\sum _{i=1}^{D} 
dy_{i}^{2}}{ |\ep |^{-4}-\sum _{i=1}^{D} y_{i}^{2} } .
\ee
Introduce a new co-ordinate $r$ by
$r^{2}=\sum _{i=1}^{D} y_{i}^{2}$
so that the metric takes the form
\be
ds^{2}=\left( r^{2}+|\ep |^{4} \right) dt^{2} +
\left( r^{2}+|\ep |^{4} \right) ^{-1} dr^{2} 
+\frac {r^{2}}{|\ep |^{4}} d\Omega _{D-2}^{2}
\ee
where $d\Omega _{D-2}^{2}$ is the metric on the $(D-2)$-sphere.
Since $|\ep |\ll 1$, the sphere has very large radius and is almost
flat, so we can introduce co-ordinates $x_{i}$ near a point on the
sphere such that
\be
|\ep |^{-4}d\Omega _{D-2}^{2} =
\sum _{i=1}^{D-2} dx_{i}^{2}.
\ee
The metric then becomes
\be
ds^{2} =\left( r^{2}+|\ep |^{4} \right) dt^{2}
+\left( r^{2}+|\ep |^{4} \right) ^{-1} dr^{2}
+r^{2} \sum _{i=1}^{D-2} dx_{i}^{2}.
\ee
We now consider a massless scalar field on this background, satisfying
the equation:
\be
\partial _{\mu }\left( {\sqrt {g}} g^{\mu \nu } 
\partial _{\nu } \Phi \right) =0,
\ee
where we assume that $\Phi $ has the form
\be
\Phi (t,r,x_{i})=f(r) \exp \left( i\sum _{i=1}^{D-2} k_{i}x_{i} 
\right) .
\ee
This gives the following equation for $f$:
\be
r^{2}\left( r^{2}+|\ep |^{4} \right) \frac {d^{2}f}{dr^{2}}
+\left[ Dr^{3}+(D-2) |\ep |^{4} r \right] \frac {df}{dr}
-k^{2} f =0.
\ee
Changing variables to $\xi =r^{2}$ gives the equation
\be
4\xi ^{2} \left( \xi +|\ep |^{4} \right) 
\frac {d^{2}f}{d\xi ^{2}} +\left[
2(D+1) \xi ^{2} +2(D-1) \xi |\ep |^{4} \right]
\frac {df}{d\xi } -k^{2}f =0.
\ee
This can be converted into the standard hypergeometric form
\be
v(1-v) \frac {d^{2}f}{dv^{2}} +
\frac {1}{2} \left[ 3-D-(5-D) v \right] \frac {df}{dv}
+ \frac {k^{2}}{4|\epsilon |^{4}}f =0,
\label{hyper}
\ee
where $v=-|\ep |^{4}/\xi $.
The equation (\ref{hyper}) has singularities at $v=0,1,\infty $,
but we are interested only in the region $v\in (-\infty ,0)$.
Therefore we shall impose the boundary conditions that $f$ is 
regular at both $v=0$ and as $v\rightarrow -\infty $.
The properties of the hypergeometric equation (\ref{hyper}) are 
well known, and reveal that for generic values of the parameters $k^{2}$
and $D$ a solution can be found which is regular at both $v=0$
and $v\rightarrow -\infty $.
For some $D$, there may be certain discrete values of $k^{2}$ for
which there is no regular solution, resulting in points which are
missing from the continuous spectrum.
However, almost all values of $k^{2}$ are eigenvalues, and 
in particular, there is no mass gap and no discrete mass 
spectrum~\cite{witten}.

\section{The Recoil Space-Time and Glueball Masses}

We are now in a position to study the generation of a mass
gap for glueballs on our recoil space time.
The reason why this space time is relevant for the QCD problem 
has been discussed in detail in \cite{em}. Basically we 
consider the interaction of Wilson loops with magnetic monopoles
of the gauge field, which are represented as D-particle defects
in a simplified (dual) picture. 
In this picture, 
the Wilson loop is viewed as a
static ``macroscopic closed string'', and the D-particle defect is assumed 
heavy so that only quantum fluctuations in its position
are taken into account and not its recoil velocity $u_i$ which is assumed 
zero. Such fluctuations induce a time-like Liouville field, $t$, 
identified with the target time, 
which 
dresses up the deformed $\sigma$-model, restoring the world-sheet
conformal symmetry, which is broken as a result of recoil~\cite{em,kogan}.

Some important remarks are in order at this point. In the approach of 
\cite{em} 
we have considered the case of $u_i=0$ 
and decoupled the time $t$ by essentially fixing it, and 
concentrating only in the spatial part
of the metric defined over constant time slices.
This time was identified, as is the case here, with the first (time-like)
Liouville field. 
There is a second Liouville field in this approach 
which was space-like,
in agreement with the fact that 
the string theory considered there was assumed to have sub-critical 
dimension~\cite{liouvilleqcd}. The result of the time fixing 
was that the spatial part of 
the space time was a Euclidean AdS ball, which, as we discussed
above, leads to no mass gap in the glueball spectrum. 

In the present approach, in  
agreement with the analysis 
in \cite{kanti}, and to be more general, 
we consider (formally) a 
generic string theory 
which may even live in its critical space-time dimension~\footnote{The 
extra dimensions, which are present  
in case one starts from a critical 
string theory and not from a sub-critical one as in \cite{em},  
play no r\^ole on the issue of the existence of a glueball mass gap, 
and eventually may be thought
of as being compactified to provide the correct regulator for QCD
in this more general approach.}.  
For us the presence of a second (space-like) Liouville field, 
although probably
correct for a QCD description as in \cite{em}, however will not 
play a crucial r\^ole in inducing a mass gap in the glueball spectrum. 
In the present approach, 
we still identify the first 
Liouville field with the target time, as in \cite{em},  
however we do not consider it completely fixed.
It is only a {\it {particular combination}} of space and time 
co-ordinates that 
is considered {\it {frozen}}. 
As we shall see, this is the crucial step, sufficient 
to yield a mass gap
in the glueball spectrum.

We commence our analysis by first giving the general expression 
for the induced space time due to the recoil of a D-particle defect
discussed in \cite{kanti}: 
\be
ds^{2}=-dt^{2}+\sum _{i=1}^{D} 2\ep \left(
\ep y_{i} + u_{i}t \right) \Theta (t) \,
dy_{i} \, dt 
+\sum _{i=1}^{D} dy_{i}^{2}.
\label{metricrecoil}
\ee
For reasons discussed above, to make contact with the QCD problem 
we take~\cite{em} the limit $u_{i}\rightarrow 0$ and consider $t>0$ only:
\be
ds^{2}=-dt^{2}+\sum _{i=1}^{D} 2\ep ^{2}  y_{i}  \,
dy_{i} \, dt +\sum _{i=1}^{D} dy_{i}^{2}.
\ee
Now introduce a new variable $r$ by
$r^{2}=\sum _{i=1}^{D} y_{i}^{2},$
in terms of which the metric (\ref{metricrecoil}) becomes
\be
ds^{2}=-dt^{2}+2\epsilon ^{2} r \, dr \, dt +dr^{2}+ r^{2}
\, d\Omega _{D-1}^{2},
\label{minkmetric}
\ee
where $d\Omega _{D-1}^{2}$ is the metric on the $D-1$ sphere.
Before Euclideanizing the metric, we first make a co-ordinate 
transformation in order to put the metric into diagonal form.
To do this, let
\be
{\tilde {t}} =t- \frac {\epsilon ^{2}}{2} r^{2}
\label{tdef}
\ee
so that the metric is now
\be
ds ^{2} = -d{\tilde {t}}^{2} 
+ \left( 1+ C r^{2} \right) dr^{2}
+r^{2} \, d\Omega _{D-1}^{2},
\ee
where 
\be
C=|\epsilon |^{4},
\label{cdef}
\ee
since $\epsilon $ is real.
The supergravity analysis of the glueball spectrum takes place 
on Euclidean space time, so now we perform a Wick rotation
${\tilde {t}}\rightarrow i{\tilde {t}}$.  
There is a subtlety associated with this procedure, namely
that we must also transform 
$\epsilon ^{-2} \rightarrow i\epsilon ^{-2}$.
This is apparent from either the non-diagonalized form of the 
metric (\ref{minkmetric}) or the definition
of ${\tilde {t}}$ (\ref{tdef}), which shows that the
Wick rotation in ${\tilde {t}}$ results from transforming both
$t$ and $\epsilon ^{-2}$.
The necessity of this manoevre is not surprising since 
$\epsilon ^{-2}$ is identified with the ``time'' 
in the Liouville string approach \cite{em}.  
The Euclidean metric then takes the form
\be
ds^{2} = d{\tilde {t}}^{2} 
+ \left( 1 - C r^{2} \right) dr^{2}
+r^{2} \, d\Omega _{D-1}^{2},
\ee
where $C=|\epsilon |^{4}$, as before.
Notice that the sign has changed in the $g_{rr}$ component of the
metric due to the transformation of $\epsilon ^{2}$.
By defining another new variable, ${\tilde {r}}$ by
\be
{\tilde {r}} = \frac {1}{2} |\epsilon |^{2} r,
\ee
we can also write the metric in the alternative form
\be
ds^{2}= d{\tilde {t}}^{2}+\frac {1}{C} 
(1-{\tilde {r}}^{2}) \, d{\tilde {r}}^2 +\frac {1}{C} 
{\tilde {r}}^{2} \, d\Omega _{D-1}^{2}.
\label{tmetricall}
\ee
The $1/C\sim 1/|\ep |^{4}$ term multiplying 
the metric on the sphere means that
the sphere has a very large radius.  
In this case we may change to Cartesian co-ordinates $x_{i}$ locally
on the sphere, which yields the metric:
\be
ds^{2}=d{\tilde {t}}^{2} +
\frac {1}{C} (1-{\tilde {r}}^{2}) \, d{\tilde {r}}^{2} +
\frac {1}{C} {\tilde {r}}^{2} \sum _{i=1}^{D-1} dx_{i}^{2}.
\label{cartmetricall}
\ee

Before we consider the massless scalar field equation, we need to 
consider the geometry of our space time.
Firstly, the original, Lorentzian geometry (\ref{metricrecoil}) has a 
$\delta $-function singularity at $t=0$.
Therefore, in order to avoid this singularity, it must be the case
that, from (\ref{tdef}), 
\be
r^{2} > -\frac {2{\tilde {t}}}{\epsilon ^{2}} .
\label{boundlower}
\ee
Therefore, for negative ${\tilde {t}}$, we have a lower bound
on the radial co-ordinate $r$.
Secondly, the spatial part of the geometry (\ref{tmetricall})
has the (Euclidean) form:
\be
ds_{S}^{2}= \left( 1-Cr^{2} \right) dr^{2}
+r^{2} \, d\Omega _{D-1}^{2}.
\label{metricbonus}
\ee
The Riemann tensor of this spatial metric has the form of a 
constant negative curvature, maximally symmetric, space to leading 
order in $\epsilon $.
This was noted previously in \cite{em} (for the original form of 
the metric (\ref{metricrecoil}), whose curvature tensor
components can be found in \cite{kanti}).
Therefore, as observed in a more complex recoil geometry in
\cite{emw}, the geometry is locally isomorphic to AdS.
However, this identification is valid only for 
\be
C r^{2} \ll 1,
\ee
which implies that
\be
r^{2} \ll  \frac {1}{C} = |\epsilon |^{-4}.
\label{upbound}
\ee
In addition, the geometry has a curvature singularity when $Cr^{2}=1$, so 
this upper bound also means that we are avoiding this second
curvature singularity.
A comment is in order concerning the magnitude of the cut-offs.
The analysis which produced the original recoil geometry 
is valid only if $|\epsilon | \ll 1$, so that the
lower bound (\ref{boundlower}) has $r$ relatively large. 
This means that we are in the outer regions of the AdS geometry.
However, since $C=|\epsilon |^{4}$, this is not incompatible with
$C r^{2} \ll 1$ being a cut-off for a rather larger value of $r$.
We shall see subsequently that it is only the ratio of the 
cut-offs which affects the glueball masses.
We finally remark that it is only the spatial
part of our metric which is AdS.
In the conventional analysis of the scalar
field equation on AdS \cite{witten}, the scalar field ansatz
is such that the scalar field depends only on the spatial 
variables (see section 2), not on the (Wick rotated) time.
Therefore the fact that the temporal part of our geometry
is not AdS does not matter. 

Following these considerations, we shall now fix
${\tilde {t}}$ to have some (negative) value, and
consider the massless scalar field equation on the background
\be
ds^{2}=
(1-Cr^{2}) \, dr^{2} + r^{2} \sum _{i=1}^{D-1} dx_{i}^{2},
\label{cartmetric}
\ee
where we have reverted to the variable $r$.
We shall impose two cut-offs to the geometry
(\ref{cartmetric}), at
\be
r=a>-\frac {2{\tilde {t}}}{|\epsilon |^{2}}
\label{cutone}
\ee
and
\be
r=b<|\epsilon |^{-4} .
\label{cuttwo}
\ee 

We now consider the equation satisfied by a massless scalar field
on the background (\ref{cartmetric}):
\be
\partial _{\mu }\left( {\sqrt {g}} g^{\mu \nu }
\partial _{\nu } \Phi \right) =0.
\label{fieldeq}
\ee
We assume that the field $\Phi $ has the form
\be
\Phi (r,x_{i})= f(r)\exp 
\left( i\sum _{i=1}^{D-1} k_{i}x_{i} \right) .
\ee
Then the differential equation satisfied by $f$ is:
\be
\frac {d^{2}f}{dr^{2}} +
\left[ \frac {D-1}{r} +\frac {Cr}{1-Cr^{2}} \right]
\frac {df}{dr} -
\frac {k^{2}}{r^{2}} \left( 1-Cr^{2} \right) f =0,
\ee
where
\be
k^{2}=\sum _{i=1}^{D-1} k_{i}^{2}.
\ee
If we now change variables to $\xi =r^{2}$, we get the equation
\be
4\xi ^{2} \left( 1-C\xi \right) \frac {d^{2}f}{d\xi ^{2}}
+2\xi \left[ D-C(D-1)\xi \right] \frac {df}{d\xi }
-k^{2} \left( 1-C\xi \right) ^{2} f =0.
\label{feqn}
\ee
This equation can be recast in the form of a standard 
Sturm-Liouville equation \cite{birkoff}
\be
\frac {d}{d\xi } \left[
\frac {\xi ^{\frac {D}{2}}}{(1-C\xi )^{\frac {1}{2}}}
\frac {df}{d\xi } \right] 
-\frac {k^{2}}{4} \xi ^{\frac {D}{2}-2} \left(
1-C\xi \right) ^{\frac {1}{2}} f =0.
\ee
In addition, we require that $f$ satisfies the boundary conditions
$f=0$ when $\xi =a^{2}$, $b^{2}$ (\ref{cutone},\ref{cuttwo}).
We can now appeal to standard theorems \cite{birkoff}, which
imply that there are an infinite number of discrete
eigenvalues $k^{2}<0$ for this problem.
In other words, we have a mass gap.

Having shown that there is a discrete mass spectrum,
we shall now proceed to find the appropriate 
eigenvalues of (\ref{feqn}).
Equation (\ref{feqn}) does not possess analytic 
solutions for general $D$.
Therefore we shall use an approximation (which 
is precisely that under which our geometry can
be regarded as AdS), which will enable us to 
produce exact results.
In equation (\ref{feqn}), 
$\xi =r^{2}$, which ranges from $\xi =a^{2}$ to
$\xi =b^{2}$,  corresponding to our two cut-offs in $r$.
However, although $\xi $ will be finite, it must
be the case that $C\xi \ll 1$ in order for our
geometry to be approximated by AdS.
In this regime equation (\ref{feqn}) reduces to
\be
4\xi ^{2} \frac {d^{2}f}{d\xi ^{2}}
+2D\xi \frac {df}{d\xi } -k^{2} f =0.
\ee
Under the variable transformation $u=\log \xi $, this
equation becomes
\be
4\frac {d^{2}f}{du^{2}} +\left( 2D-4 \right)
\frac {df}{du} -k^{2}f=0 ,
\ee
which has solutions of the form
\be
f(u) ={\cal {A}} \exp \left( \frac {(2-D)u}{4} \right)
\sin \left( \frac {\lambda (u-\delta )}{4} \right),
\ee
where ${\cal {A}}$ and $\delta $ are arbitrary constants, and 
$\lambda $  satisfies
\be
-\lambda ^{2} = 4k^{2}+(D-2)^{2} .
\label{lambdaeqn}
\ee
The solution $f$ must be sinusoidal if it is to 
vanish at the two cut-offs in $\xi $. 
Therefore it must be the case that $\lambda $ is
real, so that
\be
k^{2}< -\frac {(D-2)^{2}}{4} .
\ee
We require $f$ to vanish when $u=2\log a$, $2\log b$,
which implies that
\be
\delta = 2 \log a , \qquad
\lambda ^{2} = 4n^{2} \pi ^{2} \left[
\log  \left( \frac {b}{a} \right) \right] ^{-2} ,
\ee
where $n$ is a positive integer.
Substituting in (\ref{lambdaeqn}) gives
\be
k^{2}=-\frac {(D-2)}{4} - n^{2}\pi ^{2}
\left[ \log \left( \frac {b}{a} \right) \right] ^{-2} 
=-C_{0}-C_{1}n^{2} .
\label{evalues}
\ee
\par
The glueball masses are determined by this procedure
only up to an overall scale factor, and hence only their ratios
acquire unambiguous physical meaning.  
Furthermore, we do not at this stage know what the 
numerical values of the cut-offs $b$ and $a$ should be. 
Therefore, the analysis above predicts that the glueball masses should
depend linearly on $n^{2}$, where $n$ is a positive integer.
In the next section we shall determine the value of the 
ratio $b/a$ by demanding that our results agree with those 
of the supergravity calculation~\cite{koch}-\cite{zyskin}.

\section{Comparison with the Supergravity Approach}

It is interesting to see how the above analysis 
compares with previous numerical calculations of the
glueball masses in the supergravity approach.
As emphasized in the introduction, the important 
feature of our approach is that the region of the parameters
in which perturbation theory applies in the bulk of AdS in our framework 
is compatible with the continuum limit of the lattice QCD in which 
the glueball masses have been calculated. It is therefore interesting 
to examine whether the results obtained above are in 
agreement with the supergravity results for glueball masses
in \cite{koch}-\cite{zyskin}, which would be an independent 
confirmation of the claims~\cite{csaki} that higher-order in $\alpha'$ 
supergravity corrections in the bulk of AdS do not affect qualitatively 
the results, as far as the ratio of glueball masses are concerned. 
   
Firstly, the paper of Cs\'aki et al \cite{csaki} predicts, using
an analytic WKB approximation, that the squared glueball masses 
for the $O^{++}$ state in $QCD_{3}$ should be
proportional to $n(n+1)$.
Their numerical calculations confirm this prediction to a high order
of accuracy.
We note that our prediction of linear dependence on $n^{2}$ will be a
good approximation to $n(n+1)$ when $n$ is large.
The proportionality factor of $6$ in \cite{csaki}
was refined by further WKB analysis
in \cite{minahan}, although the correction for $QCD_{3}$ is
small and there is good agreement with the numerical values given
in \cite{csaki}. 
Minahan \cite{minahan} extends the WKB analysis to $O^{++}$
glueballs in $QCD_{4}$, where the squared masses are proportional
to $n(n+2)$.

\begin{figure}[t]
\begin{centering}
\setlength{\unitlength}{0.240900pt}
\ifx\plotpoint\undefined\newsavebox{\plotpoint}\fi
\sbox{\plotpoint}{\rule[-0.200pt]{0.400pt}{0.400pt}}%
\begin{picture}(1500,900)(0,0)
\font\gnuplot=cmr10 at 10pt
\gnuplot
\sbox{\plotpoint}{\rule[-0.200pt]{0.400pt}{0.400pt}}%
\put(176.0,113.0){\rule[-0.200pt]{303.534pt}{0.400pt}}
\put(176.0,113.0){\rule[-0.200pt]{0.400pt}{184.048pt}}
\put(176.0,113.0){\rule[-0.200pt]{4.818pt}{0.400pt}}
\put(154,113){\makebox(0,0)[r]{0}}
\put(1416.0,113.0){\rule[-0.200pt]{4.818pt}{0.400pt}}
\put(176.0,240.0){\rule[-0.200pt]{4.818pt}{0.400pt}}
\put(154,240){\makebox(0,0)[r]{200}}
\put(1416.0,240.0){\rule[-0.200pt]{4.818pt}{0.400pt}}
\put(176.0,368.0){\rule[-0.200pt]{4.818pt}{0.400pt}}
\put(154,368){\makebox(0,0)[r]{400}}
\put(1416.0,368.0){\rule[-0.200pt]{4.818pt}{0.400pt}}
\put(176.0,495.0){\rule[-0.200pt]{4.818pt}{0.400pt}}
\put(154,495){\makebox(0,0)[r]{600}}
\put(1416.0,495.0){\rule[-0.200pt]{4.818pt}{0.400pt}}
\put(176.0,622.0){\rule[-0.200pt]{4.818pt}{0.400pt}}
\put(154,622){\makebox(0,0)[r]{800}}
\put(1416.0,622.0){\rule[-0.200pt]{4.818pt}{0.400pt}}
\put(176.0,750.0){\rule[-0.200pt]{4.818pt}{0.400pt}}
\put(154,750){\makebox(0,0)[r]{1000}}
\put(1416.0,750.0){\rule[-0.200pt]{4.818pt}{0.400pt}}
\put(176.0,877.0){\rule[-0.200pt]{4.818pt}{0.400pt}}
\put(154,877){\makebox(0,0)[r]{1200}}
\put(1416.0,877.0){\rule[-0.200pt]{4.818pt}{0.400pt}}
\put(176.0,113.0){\rule[-0.200pt]{0.400pt}{4.818pt}}
\put(176,68){\makebox(0,0){0}}
\put(176.0,857.0){\rule[-0.200pt]{0.400pt}{4.818pt}}
\put(334.0,113.0){\rule[-0.200pt]{0.400pt}{4.818pt}}
\put(334,68){\makebox(0,0){20}}
\put(334.0,857.0){\rule[-0.200pt]{0.400pt}{4.818pt}}
\put(491.0,113.0){\rule[-0.200pt]{0.400pt}{4.818pt}}
\put(491,68){\makebox(0,0){40}}
\put(491.0,857.0){\rule[-0.200pt]{0.400pt}{4.818pt}}
\put(649.0,113.0){\rule[-0.200pt]{0.400pt}{4.818pt}}
\put(649,68){\makebox(0,0){60}}
\put(649.0,857.0){\rule[-0.200pt]{0.400pt}{4.818pt}}
\put(806.0,113.0){\rule[-0.200pt]{0.400pt}{4.818pt}}
\put(806,68){\makebox(0,0){80}}
\put(806.0,857.0){\rule[-0.200pt]{0.400pt}{4.818pt}}
\put(964.0,113.0){\rule[-0.200pt]{0.400pt}{4.818pt}}
\put(964,68){\makebox(0,0){100}}
\put(964.0,857.0){\rule[-0.200pt]{0.400pt}{4.818pt}}
\put(1121.0,113.0){\rule[-0.200pt]{0.400pt}{4.818pt}}
\put(1121,68){\makebox(0,0){120}}
\put(1121.0,857.0){\rule[-0.200pt]{0.400pt}{4.818pt}}
\put(1279.0,113.0){\rule[-0.200pt]{0.400pt}{4.818pt}}
\put(1279,68){\makebox(0,0){140}}
\put(1279.0,857.0){\rule[-0.200pt]{0.400pt}{4.818pt}}
\put(1436.0,113.0){\rule[-0.200pt]{0.400pt}{4.818pt}}
\put(1436,68){\makebox(0,0){160}}
\put(1436.0,857.0){\rule[-0.200pt]{0.400pt}{4.818pt}}
\put(176.0,113.0){\rule[-0.200pt]{303.534pt}{0.400pt}}
\put(1436.0,113.0){\rule[-0.200pt]{0.400pt}{184.048pt}}
\put(176.0,877.0){\rule[-0.200pt]{303.534pt}{0.400pt}}
\put(806,23){\makebox(0,0){$n^{2}$}}
\put(176.0,113.0){\rule[-0.200pt]{0.400pt}{184.048pt}}
\put(240,812){\makebox(0,0)[l]{numerical glueball 
masses squared - $QCD_3$}}
\put(220,812){\raisebox{-.8pt}{\makebox(0,0){$\circ$}}}
\put(184,120){\raisebox{-.8pt}{\makebox(0,0){$\circ$}}}
\put(208,135){\raisebox{-.8pt}{\makebox(0,0){$\circ$}}}
\put(247,157){\raisebox{-.8pt}{\makebox(0,0){$\circ$}}}
\put(302,186){\raisebox{-.8pt}{\makebox(0,0){$\circ$}}}
\put(373,223){\raisebox{-.8pt}{\makebox(0,0){$\circ$}}}
\put(460,267){\raisebox{-.8pt}{\makebox(0,0){$\circ$}}}
\put(562,318){\raisebox{-.8pt}{\makebox(0,0){$\circ$}}}
\put(680,376){\raisebox{-.8pt}{\makebox(0,0){$\circ$}}}
\put(814,442){\raisebox{-.8pt}{\makebox(0,0){$\circ$}}}
\put(964,515){\raisebox{-.8pt}{\makebox(0,0){$\circ$}}}
\put(1129,596){\raisebox{-.8pt}{\makebox(0,0){$\circ$}}}
\put(1310,683){\raisebox{-.8pt}{\makebox(0,0){$\circ$}}}
\put(300,767){\makebox(0,0)[l]{best fit line}}
\multiput(220,767)(20.756,0.000){4}{\usebox{\plotpoint}}
\put(184,126){\usebox{\plotpoint}}
\put(184.00,126.00){\usebox{\plotpoint}}
\put(202.36,135.68){\usebox{\plotpoint}}
\multiput(207,138)(18.895,8.589){0}{\usebox{\plotpoint}}
\put(221.06,144.67){\usebox{\plotpoint}}
\put(239.47,154.24){\usebox{\plotpoint}}
\multiput(241,155)(18.895,8.589){0}{\usebox{\plotpoint}}
\put(258.11,163.33){\usebox{\plotpoint}}
\multiput(263,166)(18.564,9.282){0}{\usebox{\plotpoint}}
\put(276.61,172.73){\usebox{\plotpoint}}
\put(295.34,181.67){\usebox{\plotpoint}}
\multiput(298,183)(18.221,9.939){0}{\usebox{\plotpoint}}
\put(313.78,191.17){\usebox{\plotpoint}}
\multiput(320,194)(18.564,9.282){0}{\usebox{\plotpoint}}
\put(332.45,200.24){\usebox{\plotpoint}}
\put(350.95,209.61){\usebox{\plotpoint}}
\multiput(354,211)(18.564,9.282){0}{\usebox{\plotpoint}}
\put(369.50,218.91){\usebox{\plotpoint}}
\put(388.28,227.70){\usebox{\plotpoint}}
\multiput(389,228)(18.221,9.939){0}{\usebox{\plotpoint}}
\put(406.53,237.56){\usebox{\plotpoint}}
\multiput(411,240)(19.159,7.983){0}{\usebox{\plotpoint}}
\put(425.34,246.28){\usebox{\plotpoint}}
\put(443.56,256.22){\usebox{\plotpoint}}
\multiput(445,257)(19.159,7.983){0}{\usebox{\plotpoint}}
\put(462.37,264.93){\usebox{\plotpoint}}
\multiput(468,268)(18.564,9.282){0}{\usebox{\plotpoint}}
\put(480.85,274.38){\usebox{\plotpoint}}
\put(499.43,283.60){\usebox{\plotpoint}}
\multiput(502,285)(18.564,9.282){0}{\usebox{\plotpoint}}
\put(518.01,292.82){\usebox{\plotpoint}}
\put(536.70,301.85){\usebox{\plotpoint}}
\multiput(537,302)(18.221,9.939){0}{\usebox{\plotpoint}}
\put(555.18,311.27){\usebox{\plotpoint}}
\multiput(559,313)(18.564,9.282){0}{\usebox{\plotpoint}}
\put(573.76,320.51){\usebox{\plotpoint}}
\put(592.35,329.71){\usebox{\plotpoint}}
\multiput(593,330)(18.564,9.282){0}{\usebox{\plotpoint}}
\put(610.82,339.17){\usebox{\plotpoint}}
\multiput(616,342)(18.895,8.589){0}{\usebox{\plotpoint}}
\put(629.48,348.24){\usebox{\plotpoint}}
\put(647.88,357.84){\usebox{\plotpoint}}
\multiput(650,359)(19.159,7.983){0}{\usebox{\plotpoint}}
\put(666.68,366.56){\usebox{\plotpoint}}
\multiput(673,370)(18.221,9.939){0}{\usebox{\plotpoint}}
\put(684.95,376.40){\usebox{\plotpoint}}
\put(703.71,385.21){\usebox{\plotpoint}}
\multiput(707,387)(18.221,9.939){0}{\usebox{\plotpoint}}
\put(722.14,394.72){\usebox{\plotpoint}}
\put(740.74,403.86){\usebox{\plotpoint}}
\multiput(741,404)(18.564,9.282){0}{\usebox{\plotpoint}}
\put(759.42,412.92){\usebox{\plotpoint}}
\multiput(764,415)(18.221,9.939){0}{\usebox{\plotpoint}}
\put(777.85,422.43){\usebox{\plotpoint}}
\put(796.59,431.36){\usebox{\plotpoint}}
\multiput(798,432)(18.221,9.939){0}{\usebox{\plotpoint}}
\put(814.97,440.98){\usebox{\plotpoint}}
\multiput(821,444)(18.895,8.589){0}{\usebox{\plotpoint}}
\put(833.72,449.86){\usebox{\plotpoint}}
\put(852.14,459.44){\usebox{\plotpoint}}
\multiput(855,461)(18.895,8.589){0}{\usebox{\plotpoint}}
\put(870.84,468.42){\usebox{\plotpoint}}
\multiput(878,472)(18.221,9.939){0}{\usebox{\plotpoint}}
\put(889.20,478.08){\usebox{\plotpoint}}
\put(908.00,486.82){\usebox{\plotpoint}}
\multiput(912,489)(18.221,9.939){0}{\usebox{\plotpoint}}
\put(926.39,496.41){\usebox{\plotpoint}}
\put(945.03,505.47){\usebox{\plotpoint}}
\multiput(946,506)(18.221,9.939){0}{\usebox{\plotpoint}}
\put(963.57,514.74){\usebox{\plotpoint}}
\multiput(969,517)(18.221,9.939){0}{\usebox{\plotpoint}}
\put(982.10,524.05){\usebox{\plotpoint}}
\put(1000.82,533.01){\usebox{\plotpoint}}
\multiput(1003,534)(18.221,9.939){0}{\usebox{\plotpoint}}
\put(1019.21,542.61){\usebox{\plotpoint}}
\multiput(1026,546)(18.895,8.589){0}{\usebox{\plotpoint}}
\put(1037.95,551.52){\usebox{\plotpoint}}
\put(1056.33,561.16){\usebox{\plotpoint}}
\multiput(1060,563)(18.895,8.589){0}{\usebox{\plotpoint}}
\put(1075.08,570.04){\usebox{\plotpoint}}
\put(1093.45,579.70){\usebox{\plotpoint}}
\multiput(1094,580)(18.895,8.589){0}{\usebox{\plotpoint}}
\put(1112.20,588.60){\usebox{\plotpoint}}
\multiput(1117,591)(18.221,9.939){0}{\usebox{\plotpoint}}
\put(1130.60,598.18){\usebox{\plotpoint}}
\put(1149.31,607.16){\usebox{\plotpoint}}
\multiput(1151,608)(18.221,9.939){0}{\usebox{\plotpoint}}
\put(1167.77,616.62){\usebox{\plotpoint}}
\multiput(1173,619)(18.564,9.282){0}{\usebox{\plotpoint}}
\put(1186.40,625.76){\usebox{\plotpoint}}
\put(1205.06,634.78){\usebox{\plotpoint}}
\multiput(1208,636)(18.221,9.939){0}{\usebox{\plotpoint}}
\put(1223.43,644.42){\usebox{\plotpoint}}
\multiput(1230,648)(19.159,7.983){0}{\usebox{\plotpoint}}
\put(1242.24,653.13){\usebox{\plotpoint}}
\put(1260.60,662.80){\usebox{\plotpoint}}
\multiput(1265,665)(18.895,8.589){0}{\usebox{\plotpoint}}
\put(1279.29,671.80){\usebox{\plotpoint}}
\put(1297.71,681.36){\usebox{\plotpoint}}
\multiput(1299,682)(18.895,8.589){0}{\usebox{\plotpoint}}
\put(1310,687){\usebox{\plotpoint}}
\sbox{\plotpoint}{\rule[-0.400pt]{0.800pt}{0.800pt}}%
\put(240,722){\makebox(0,0)[l]{numerical glueball masses 
squared - $QCD_4$}}
\put(220,722){\makebox(0,0){$+$}}
\put(184,130){\makebox(0,0){$+$}}
\put(208,154){\makebox(0,0){$+$}}
\put(247,186){\makebox(0,0){$+$}}
\put(302,226){\makebox(0,0){$+$}}
\put(373,275){\makebox(0,0){$+$}}
\put(460,333){\makebox(0,0){$+$}}
\put(562,399){\makebox(0,0){$+$}}
\put(680,473){\makebox(0,0){$+$}}
\put(814,556){\makebox(0,0){$+$}}
\put(964,648){\makebox(0,0){$+$}}
\put(1129,748){\makebox(0,0){$+$}}
\sbox{\plotpoint}{\rule[-0.500pt]{1.000pt}{1.000pt}}%
\put(300,677){\makebox(0,0)[l]{best fit line}}
\multiput(220,677)(20.756,0.000){4}{\usebox{\plotpoint}}
\put(184,146){\usebox{\plotpoint}}
\put(184.00,146.00){\usebox{\plotpoint}}
\put(201.67,156.89){\usebox{\plotpoint}}
\multiput(207,160)(16.786,12.208){0}{\usebox{\plotpoint}}
\put(218.83,168.53){\usebox{\plotpoint}}
\put(236.51,179.38){\usebox{\plotpoint}}
\multiput(241,182)(16.786,12.208){0}{\usebox{\plotpoint}}
\put(253.65,191.05){\usebox{\plotpoint}}
\put(271.05,202.37){\usebox{\plotpoint}}
\multiput(275,205)(17.511,11.143){0}{\usebox{\plotpoint}}
\put(288.56,213.50){\usebox{\plotpoint}}
\put(305.95,224.78){\usebox{\plotpoint}}
\multiput(309,227)(17.511,11.143){0}{\usebox{\plotpoint}}
\put(323.41,235.99){\usebox{\plotpoint}}
\put(340.74,247.36){\usebox{\plotpoint}}
\multiput(343,249)(17.511,11.143){0}{\usebox{\plotpoint}}
\put(358.10,258.73){\usebox{\plotpoint}}
\put(375.50,270.05){\usebox{\plotpoint}}
\multiput(377,271)(17.928,10.458){0}{\usebox{\plotpoint}}
\put(393.11,280.99){\usebox{\plotpoint}}
\put(410.33,292.57){\usebox{\plotpoint}}
\multiput(411,293)(17.928,10.458){0}{\usebox{\plotpoint}}
\put(427.90,303.57){\usebox{\plotpoint}}
\multiput(434,308)(17.511,11.143){0}{\usebox{\plotpoint}}
\put(445.15,315.10){\usebox{\plotpoint}}
\put(462.49,326.50){\usebox{\plotpoint}}
\multiput(468,330)(17.928,10.458){0}{\usebox{\plotpoint}}
\put(480.27,337.20){\usebox{\plotpoint}}
\put(497.32,349.02){\usebox{\plotpoint}}
\multiput(502,352)(17.928,10.458){0}{\usebox{\plotpoint}}
\put(515.06,359.77){\usebox{\plotpoint}}
\put(532.32,371.27){\usebox{\plotpoint}}
\multiput(537,374)(16.786,12.208){0}{\usebox{\plotpoint}}
\put(549.46,382.93){\usebox{\plotpoint}}
\put(567.16,393.76){\usebox{\plotpoint}}
\multiput(571,396)(16.786,12.208){0}{\usebox{\plotpoint}}
\put(584.29,405.45){\usebox{\plotpoint}}
\put(602.01,416.25){\usebox{\plotpoint}}
\multiput(605,418)(16.786,12.208){0}{\usebox{\plotpoint}}
\put(619.11,427.98){\usebox{\plotpoint}}
\put(636.49,439.33){\usebox{\plotpoint}}
\multiput(639,441)(17.511,11.143){0}{\usebox{\plotpoint}}
\put(654.06,450.37){\usebox{\plotpoint}}
\put(671.35,461.80){\usebox{\plotpoint}}
\multiput(673,463)(17.511,11.143){0}{\usebox{\plotpoint}}
\put(688.91,472.86){\usebox{\plotpoint}}
\put(706.14,484.38){\usebox{\plotpoint}}
\multiput(707,485)(17.511,11.143){0}{\usebox{\plotpoint}}
\put(723.54,495.69){\usebox{\plotpoint}}
\put(740.96,506.97){\usebox{\plotpoint}}
\multiput(741,507)(17.928,10.458){0}{\usebox{\plotpoint}}
\put(758.51,518.01){\usebox{\plotpoint}}
\multiput(764,522)(17.511,11.143){0}{\usebox{\plotpoint}}
\put(775.80,529.47){\usebox{\plotpoint}}
\put(793.30,540.58){\usebox{\plotpoint}}
\multiput(798,544)(17.511,11.143){0}{\usebox{\plotpoint}}
\put(810.59,552.06){\usebox{\plotpoint}}
\put(827.95,563.43){\usebox{\plotpoint}}
\multiput(832,566)(17.928,10.458){0}{\usebox{\plotpoint}}
\put(845.67,574.22){\usebox{\plotpoint}}
\put(862.78,585.95){\usebox{\plotpoint}}
\multiput(866,588)(17.928,10.458){0}{\usebox{\plotpoint}}
\put(880.46,596.79){\usebox{\plotpoint}}
\put(897.81,608.14){\usebox{\plotpoint}}
\multiput(901,610)(16.786,12.208){0}{\usebox{\plotpoint}}
\put(914.92,619.86){\usebox{\plotpoint}}
\put(932.66,630.63){\usebox{\plotpoint}}
\multiput(935,632)(16.786,12.208){0}{\usebox{\plotpoint}}
\put(949.75,642.38){\usebox{\plotpoint}}
\put(967.50,653.13){\usebox{\plotpoint}}
\multiput(969,654)(16.786,12.208){0}{\usebox{\plotpoint}}
\put(984.68,664.73){\usebox{\plotpoint}}
\put(1001.93,676.22){\usebox{\plotpoint}}
\multiput(1003,677)(17.511,11.143){0}{\usebox{\plotpoint}}
\put(1019.53,687.22){\usebox{\plotpoint}}
\put(1036.73,698.80){\usebox{\plotpoint}}
\multiput(1037,699)(17.511,11.143){0}{\usebox{\plotpoint}}
\put(1054.37,709.72){\usebox{\plotpoint}}
\multiput(1060,713)(16.786,12.208){0}{\usebox{\plotpoint}}
\put(1071.55,721.32){\usebox{\plotpoint}}
\put(1089.07,732.41){\usebox{\plotpoint}}
\multiput(1094,736)(17.511,11.143){0}{\usebox{\plotpoint}}
\put(1106.40,743.81){\usebox{\plotpoint}}
\put(1123.86,754.99){\usebox{\plotpoint}}
\multiput(1128,758)(17.511,11.143){0}{\usebox{\plotpoint}}
\put(1141.24,766.31){\usebox{\plotpoint}}
\put(1158.65,777.56){\usebox{\plotpoint}}
\multiput(1162,780)(17.511,11.143){0}{\usebox{\plotpoint}}
\put(1176.09,788.80){\usebox{\plotpoint}}
\put(1193.44,800.14){\usebox{\plotpoint}}
\multiput(1196,802)(17.928,10.458){0}{\usebox{\plotpoint}}
\put(1210.99,811.18){\usebox{\plotpoint}}
\put(1228.16,822.83){\usebox{\plotpoint}}
\multiput(1230,824)(17.928,10.458){0}{\usebox{\plotpoint}}
\put(1245.78,833.75){\usebox{\plotpoint}}
\put(1263.22,844.96){\usebox{\plotpoint}}
\multiput(1265,846)(17.511,11.143){0}{\usebox{\plotpoint}}
\put(1280.57,856.33){\usebox{\plotpoint}}
\put(1298.07,867.46){\usebox{\plotpoint}}
\multiput(1299,868)(16.786,12.208){0}{\usebox{\plotpoint}}
\put(1310,876){\usebox{\plotpoint}}
\end{picture}
\caption{A comparison of the exact glueball masses squared calculated
numerically in supergravity \cite{koch} and the predicted
linear dependence on $n^{2}$.
Values for the first twelve $O^{++}$ glueball states are used in
both $QCD_{3}$ and $QCD_{4}$.}
\label{figure}
\end{centering}
\end{figure}
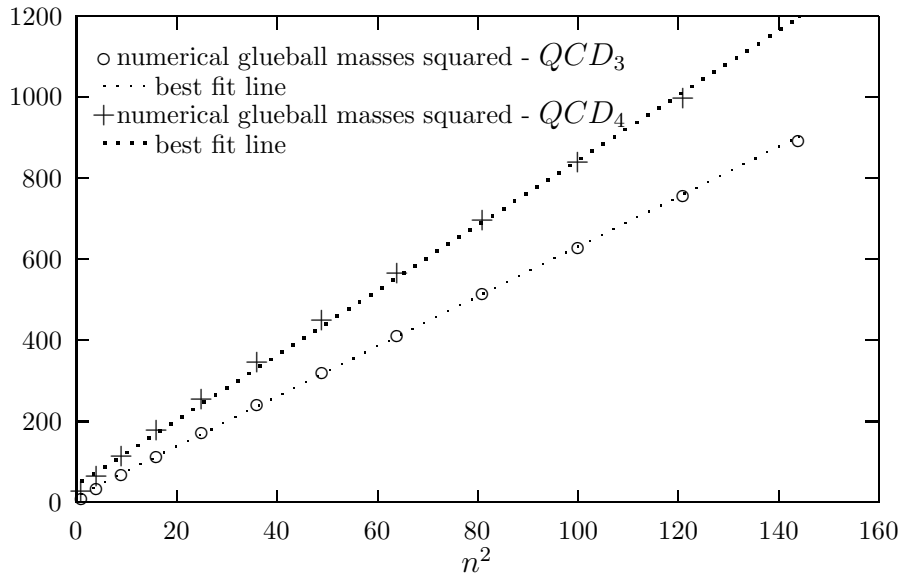

However, here we consider the numerical values calculated by de Mello
Koch et al \cite{koch}, since they give the glueball masses for
much higher values of $n$ than other authors (we note that 
the authors of \cite{koch} are sceptical about the existence
of an exact mass formula).  
We plot in figure \ref{figure} the squared
masses for the first twelve glueball states given in \cite{koch}
for the $O^{++}$ glueball in $QCD_{3}$ and
$QCD_{4}$, against $n^{2}$.
We also show a best fit linear regression line in each case.
It can be seen that the linear dependence on $n^{2}$ is a very good
approximation to these glueball masses.
Since there is freedom in the overall scale of the squared
masses, only the ratio of the coefficients in (\ref{evalues})
is relevant physically, and this determines (or, is determined by)
the ratio of the cut-offs, once the dimension of the geometry
is fixed.

The (numerical) results 
of the best fits are given below, and in figure \ref{figure}.  
We observe from the best fit lines in figure \ref{figure}
that there is agreement between our approach and 
the glueball masses derived from supergravity \cite{koch},
provided that the ratio of the two cut-offs in our approach
(\ref{cutone},\ref{cuttwo}) is fixed to be of order 3.
To see this, note that the best fit lines for $QCD_{3}$ 
and $QCD_{4}$ are given by, respectively,
\be
14.73+6.16 n^{2}
\ee
and
\be
43.12 + 8.02 n^{2}
\ee
The ratios of the coefficients in (\ref{evalues}) can then be
used to predict the ratio of the cut-offs, $b/a$, once
we have specified the number of dimensions, $D$.
For $QCD_{3}$, it is appropriate to consider 5-dimensional AdS,
so we set $D=5$, whilst for $QCD_{4}$, we need $D=7$ \cite{csaki}.
The numerical values for $QCD_{3}$ lead to 
$\log (b/a) = 3.2$, whilst those for $QCD_{4}$ give
$\log (b/a) = 2.9$.

We find it striking that a single number for the
two scales in the target space geometry suffices to describe
consistently glueball masses in QCD in various dimensions. 
This agreement may be taken as a confirmation of the claims~\cite{csaki}
that the results of the (ratios of) 
glueball masses obtained using the lowest-order
large-N limit in the supergravity approach of \cite{maldacena,witten}
to QCD, survive higher-order $\alpha'$ corrections, as more appropriate
in that approach for comparison with the continuum limit of the 
lattice results for the 
glueball masses.

\section{Conclusions}

In this article 
we have discussed in detail the geometric properties of
a space time induced {\it dynamically} as a result of the recoil 
of a heavy D-particle defect during scattering with a macroscopic closed 
string loop, in the limit of vanishing recoil veolcity $u_i \rightarrow 0$. 
As emphasized in \cite{em}, such space times may be relevant for 
understanding nonperturbative infrared properties of QCD-like gauge theories,
within the framework of Liouville strings~\cite{liouvilleqcd}.  

In the present article we have derived the properties of the 
D-particle-recoil induced space time pertinent to 
the mass gap, and compared
with the case of AdS (both with, and without an event
horizon).
The large mass black hole
space time (\ref{wittenmetric}) has a co-ordinate singularity
at the horizon $\rho = b$, and a genuine singularity at $\rho =0$, 
for example:
\be
R_{\mu \nu \sigma \lambda }R^{\mu \nu \sigma \lambda }
=\frac {8(5\rho ^{8}+9b^{8})}{b^{4}\rho ^{8}} .
\ee  
According to Witten \cite{witten}, the presence of a ``cut-off'' in
the metric (\ref{wittenmetric}) is crucial for the mass gap.
However, performing a similar calculation in AdS with a cut-off 
(i.e. the same as in section 2, but with a cut-off at some value of
$v=v_{0}\in (-\infty ,0)$) does not give a mass gap, since the value
of $v$ at the cut-off will be a regular point of the equation
(\ref{hyper}).
Therefore, a cut-off is not, on its own, sufficient to generate a
mass gap.
This means that the co-ordinate singularity in (\ref{wittenmetric})
must be crucial.

The recoil space time has a genuine singularity when 
$t=0$, which leads to the imposition of a lower cut-off on $r$,
at $r=a$.
Note that it is not possible to set $a=0$ and get the same result.
The point $\xi =0$ is a regular singular point of the equation
(\ref{feqn}), and a series solution about this point can be found
by the usual Frobenius method. 
In order to have $f(0)=0$, it must be the case that $k^{2}>0$,
which is not what we want for the mass gap.
Therefore, it is necessary to impose a positive cut-off on $r$, no
matter how small. 
By choosing $\epsilon $ sufficiently small, it is possible to take
the upper cut-off value $b$ as large as we like. 
This means that the argument of Witten \cite{witten}, which led to the
interpretation of the mass gap in terms of glueball masses on the
boundary Minkowski space, is still valid for our
situation since we can cover as large an area of AdS as we like (in 
particular, we can go as near as we like to infinity).

It is interesting to continue the computation of higher-tensor 
``matter'' fields in our approach and compare with higher-order 
glueball calculations in the QCD and supergravity approaches, 
as well as to probe 
our analysis further towards an understanding 
of the nature of the confinement-deconfinement phase transition.
Moreover, it would be interesting to place our 
``recoil'' approach in a wider
context, and to compare it with other approaches to QCD using 
Liouville-strings, such as the holographic 
renormalization-group flow~\cite{others}, 
especially in connection with the 
infrared running of the gauge coupling, which in our approach has been 
discussed briefly in \cite{em}.  
These are left for future work. 

\section*{Acknolwedgements}

It is a pleasure to thank John Ellis for discussions.
The work of N.E.M. is partially supported by P.P.A.R.C. (U.K.).
That of E.W. is supported by Oriel College (Oxford), and
she would like to thank the Department of Physics, University
of Newcastle (U.K.), for hospitality during the early stages of this work.

\end{document}